\magnification=1200
\hfuzz=3pt
\hsize=12.5cm
\hoffset=0.32cm
\baselineskip=18pt
\voffset=\baselineskip
\nopagenumbers
\font\title=cmssdc10 at 20pt
\font\smalltitle=cmssdc10 at 14pt

\let\text=\textstyle

\def\RE{\mathop{\Re e}\nolimits}
\def\IM{\mathop{\Im m}\nolimits}
\def\sectionstyle{\smalltitle}
\newskip\beforesectionskip
\newskip\aftersectionskip
\beforesectionskip=4mm plus 1mm minus 1mm
\aftersectionskip=2mm plus .2mm minus .2mm
\newcount\mysectioncounter
\def\resetsections{\mysectioncounter=0}
\resetsections
\newcount\myeqcounter
\def\mysection#1\par{\par\removelastskip\penalty -250
\vskip\beforesectionskip
\global\advance\mysectioncounter by 1\noindent
\myeqcounter=0{\sectionstyle\the\mysectioncounter.
#1}\par
\nobreak\vskip\aftersectionskip}
\def\myeqno{\global\advance\myeqcounter by 1\eqno{(\the\mysectioncounter.\the\myeqcounter)}}
\def\mydispeqno{\global\advance\myeqcounter by 1\hfill\llap{(\the\mysectioncounter
.\the\myeqcounter)}}
\headline={\hfil\tenrm\folio\hfil}
\footline={\hfil}
\baselineskip=12pt
\parindent= 20pt
\centerline{\smalltitle Anomalous diffusion of a particle in an aging medium}
\vskip 0.5cm
\centerline{No\"elle POTTIER}
\centerline{\sl F\'{e}d\'{e}ration de Recherche CNRS 2438 ``Mati\`{e}re et Syst\`{e}mes Complexes''\/} 
\centerline{\sl and\/} 
\centerline{\sl Groupe de Physique des Solides, CNRS UMR 7588, Universit\'{e}s Paris 6 et Paris 7,\/} 
\centerline{\sl 2, place Jussieu, 75251 Paris Cedex 05, France\/}
\medskip
\centerline{and}
\medskip
\centerline{Alain MAUGER}
\centerline{\sl Laboratoire des Milieux D\'esordonn\'es et
H\'et\'erog\`enes, CNRS UMR 7603, Universit\'e Paris 6,\/}
\centerline{\sl Tour 22, Case 86, 4 place Jussieu, 75252 Paris Cedex 05, France.\/}
\vskip 2cm
\noindent
{\smalltitle Abstract}
\medskip
\noindent
We report new results about the anomalous diffusion of a particle in an aging medium. For each given age, the quasi-stationary
particle velocity is governed by a generalized Langevin equation with a frequency-dependent friction coefficient 
proportional to $|\omega|^{\delta-1}$ at small frequencies, with $0<\delta<2$. The aging properties of the medium are encoded
in a frequency dependent effective temperature $T_{\rm eff.}(\omega)$. The latter is modelized by a function
proportional to $|\omega|^\alpha$ at small frequencies, with $\alpha<0$, thus allowing for the medium to have a density of slow
modes proportionally larger than in a thermal bath. Using asymptotic Fourier analysis, we obtain the
behaviour at large times of the velocity correlation function and of the mean square displacement. As a result, the anomalous
diffusion exponent in the aging medium appears to be linked, not only to $\delta$ as it would be the case in a thermal bath,
but also to the exponent $\alpha$ characterizing the density of slow modes.  
\bigskip
\parindent=0pt
{\bf PACS numbers:} 

05.40.-a Fluctuation phenomena, random processes, noise and Brownian motion

02.50.Ey Stochastic processes
\vskip 1cm
{\bf KEYWORDS:} 
\medskip
{\parskip=0pt
{\sl Corresponding author\/:
\smallskip
No\"elle POTTIER
 
Fax number\/: + 33 1 46 33 94 01 

E-mail\/: npottier@ccr.jussieu.fr\/}}

\vfill
\break
\parskip=0pt
\parindent=20pt
\baselineskip=16pt
\mysection{Introduction}

The question we want to address in the present paper is that of the diffusion (possibly anomalous) of a particle 
in an out of equilibrium environment. Generally speaking, the medium in which a diffusing particle evolves may be, or
not, in a state of thermodynamic equilibrium. For instance, when it is constituted by an aging medium such as a glassy
colloidal suspension of Laponite [1]-[3], the environment of a diffusing particle is not in thermal equilibrium. This renders
the analysis of the particle motion much more involved, as compared with the thermal equilibrium case. 

As well-known, the motion of a diffusing particle in a stationary medium is properly described by a
generalized Langevin equation, in which the effect of the environment is encoded in a friction term and a noise term. When the
environment is a thermal bath, the noise spectral density and the real part of the friction coefficient are related by the
fluctuation-dissipation theorem of the second kind [4],[5]. Thermalizing particle variables such as the particle velocity then
satisfy the fluctuation-dissipation theorem of the first kind {\sl i.e.\/} the Einstein relation [4],[5]. In both theorems the
temperature of the bath plays the central role. 

None of these theorems is valid when the environment ages, since then it is out of
equilibrium. It has been proposed to associate to an aging system a violation factor of the fluctuation-dissipation
theorem, or an effective temperature [6],[7]. This latter quantity is considered as a function $T_{\rm
eff.}(\omega,t_w)$ of the angular frequency~$\omega$ of an external perturbation and of the age $t_w$ of the system. 

In the present paper we analyze in details how this concept allows for the description of diffusion in an aging medium. 
We propose to modelize $T_{\rm eff.}(\omega,t_w)$ by an inverse power-law of $\omega$ at small $\omega$, in
accordance with the fact that, in an aging medium, the density of slow modes is proportionally enhanced as compared to what it
is in a thermal bath (for the same real part of the friction coefficient). We show that this modelization leads to an anomalous
diffusion exponent which depends, not only on the exponent characterizing the real part of the friction coefficient, as it is
the case at equilibrium, but also on the exponent characterizing $T_{\rm eff.}$. This constitutes the main new result
of the paper.

The paper is organized as follows. In Section 2, we briefly recall the main features of the diffusion of a particle in a
stationary medium. In particular, we carefully separate out the properties valid at thermal equilibrium (namely, the
two fluctuation-dissipation theorems and the regression theorem) from those which only require stationarity. In Section 3, we
consider the general situation of a particle diffusing in an out of equilibrium environment. Assuming a bilinear coupling
between the particle and the environment, we derive an equation of motion for the particle.
This equation involves an environment response function and a fluctuating force. In contrast to the
equilibrium case, there is no specific relation between these two quantities. In Section 4, we show on general grounds
how, when there exists two well separated time scales respectively characterizing the times pertinent for the measuring
processes and the waiting time or the age of the system, a quasi-stationary regime can be defined, as pictured by
frequency-dependent generalized susceptibilities (parametrized by $t_w$). We then focus the study on the environment response
function and on the particle velocity response function. We show how, in the quasi-stationary regime which takes place when
$\omega t_w\gg 1$, an effective temperature $T_{\rm eff.}(\omega,t_w)$ can be defined, which modifies in a consistent way
both the Einstein relation and the relation between the noise spectral density and the real part of the friction coefficient.
In Section 5, we propose a modelization of $T_{\rm eff.}$ by an inverse power-law of $\omega$ at small $\omega$, which
corresponds to an enhanced density of slow modes with respect to equilibrium. Using asymptotic Fourier analysis, we compute the
anomalous diffusion exponent. As a  result, it depends on both exponents
characterizing the real part of the friction coefficient and the effective temperature. This result shows that independent
measurements of the real part of the particle mobility and of the mean square displacement will in turn allow for the
determination of the exponent associated with the small $\omega$ behaviour of $T_{\rm eff.}$, which describes the density of
slow modes in the aging medium. Finally, Section 6 contains our conclusions.

\mysection {Diffusion in a stationary medium}

To begin with, we recall in this Section some known results about the diffusion (possibly anomalous) of a particle in a
stationary medium. For further purpose, we will separate out the results valid for any stationary medium from those which are
only valid for a thermal bath.

The motion of a diffusing particle of mass $m$ evolving in a stationary medium is usually described by the generalized
Langevin equation [4],[5],
$$m\,{dv\over dt}=-m\int_{-\infty}^\infty\tilde\gamma(t-t')\,v(t')\,dt'+F(t),\qquad v={dx\over dt},\myeqno$$
in which $F(t)$ is the Langevin random force acting on the particle and $\tilde\gamma(t)$ is a retarded friction kernel.
In Eq. (2.1) it is assumed that the diffusing particle and the surrounding medium have been put in contact in an infinitely
remote past, as pictured by the lower integration bound $-\infty$ in the retarded friction term. Both $F(t)$ and the solution
$v(t)$ of the generalized Langevin equation (2.1) can be viewed as stationary random processes. Their spectral densities are
linked by
$$C_{vv}(\omega)=|\mu(\omega)|^2\,C_{FF}(\omega),\myeqno$$
where
$$\mu(\omega)={1\over m[\gamma(\omega)-i\omega]}\myeqno$$
denotes the frequency-dependent particle mobility ($\gamma(\omega)$ is the Fourier transform of $\tilde\gamma(t)$, as defined
by $\gamma(\omega)=\int_{-\infty}^\infty\tilde\gamma(t)\,e^{i\omega t}\,dt$). 
\bigskip
{\bf 2.1. Relaxation of the average velocity}

As shown by the expression (2.3) of the mobility, the function $\gamma(\omega)$ characterizes the relaxation of the average
particle velocity. For instance, with a power-law behaviour of $\gamma(\omega)$
(namely $\RE\gamma(\omega)\propto|\omega|^{\delta-1}$ with $0<\delta<2$), one has [8]:
$$\bigl\langle v(t)\bigr\rangle=v(t=0)\,E_{2-\delta}\bigl[-(\omega_\delta t)^{2-\delta}\bigr].\myeqno$$
In Eq. (2.4), $E_\alpha(x)$ denotes the Mittag-Leffler function\footnote{$^1$}{The Mittag-Leffler function is defined by the
series expansion
$$E_\alpha(x)=\sum_{n=0}^\infty{x^n\over\Gamma(\alpha n+1)},\qquad\alpha>0,$$
where $\Gamma$ is the Euler Gamma function. The Mittag-Leffler function $E_\alpha(x)$ reduces to the exponential $e^x$ when
$\alpha=1$. The asymptotic behaviour at large $x$ of the Mittag-Leffler function $E_\alpha(x)$ is as follows:
$$E_\alpha(x)\simeq-{1\over x}\,{1\over\Gamma(1-\alpha)},\qquad x\gg 1.$$} of index $\alpha$ [9],[10], and the
$\delta$-dependent frequency $\omega_\delta$
acts as an inverse relaxation time. For $\delta=1$,
the function
$E_{2-\delta}\bigl[-(\omega_\delta t)^{2-\delta}\bigr]$ reduces to a decreasing exponential
$(E_1(-\omega_1t)=e^{-\omega_1t})$, while, for all other values of
$\delta$, it decays algebraically at large times ($E_{2-\delta}\bigl[-(\omega_\delta t)^{\delta-2}\bigr]\simeq{(\omega_\delta
t)^{\delta-2}/\Gamma(\delta-1)}$) [10]. 
\bigskip
{\bf 2.2. Velocity correlation function and mean square displacement} 

Applying the Wiener-Khintchine theorem, one obtains the velocity correlation function as the inverse Fourier transform of
$C_{vv}(\omega)$, that is, in terms of the noise spectral density $C_{FF}(\omega)$:
$$\bigl\langle v(t)v(t')\bigr\rangle={1\over m^2}\,\int_{-\infty}^\infty{d\omega\over 2\pi}\,e^{-i\omega(t-t')}
\,C_{FF}(\omega)\,{1\over\gamma(\omega)-i\omega}\,{1\over\gamma^*(\omega)+i\omega}.\myeqno$$
Eq. (2.5) can be rewritten as:
$$\bigl\langle v(t)v(t')\bigr\rangle={1\over m^2}\,\int_{-\infty}^\infty{d\omega\over 2\pi}\,e^{-i\omega(t-t')}
{C_{FF}(\omega)\over
2\RE\gamma(\omega)}\,\left\{{1\over\gamma(\omega)-i\omega}+{1\over\gamma^*(\omega)+i\omega}\right\}.\myeqno$$
From the velocity correlation function, one can compute the mean square displacement of the diffusing particle (a quantity
which cannot be written under the form of a Fourier integral) as a double integral over time:
$$\Delta x^2(t)=\bigl\langle [x(t)-x(0)]^2\bigr\rangle=2\int_0^t dt_1\int_0^{t_1}dt_2\,\bigl\langle
v(t_1)v(t_2)\bigr\rangle.\myeqno$$
\bigskip
{\bf 2.3. Case of a thermal bath: fluctuation-dissipation theorems}

Let us now consider the particular case of diffusion in a thermal bath. The linear response theory, applied to the particle
velocity considered as a dynamical variable of the isolated particle-plus-bath system, allows to express the mobility in
terms of the equilibrium velocity correlation function. Since the mobility $\mu(\omega)$ is nothing but the generalized
susceptibility $\chi_{vx}(\omega)$, one has 
$$\mu(\omega)={1\over kT}\int_0^\infty\bigl\langle v(t)v\bigr\rangle\,e^{i\omega t}\,dt,\myeqno$$
where $T$ is the bath temperature [4],[5]. The result (2.8) constitutes the first fluctuation-dissipation theorem (FDT).
Accordingly, the velocity spectral density is given by:
$$C_{vv}(\omega)=\int_{-\infty}^\infty\bigl\langle v(t)v\bigr\rangle\,e^{i\omega t}\,dt=kT\,2\RE\mu(\omega).\myeqno$$
Introducing the frequency-dependent diffusion coefficient $D(\omega)$ as defined by
$$D(\omega)=\int_0^\infty\bigl\langle v(t)v\bigr\rangle\,\cos\omega t\,dt,\myeqno$$
one can rewrite Eq. (2.9) as the well-known Einstein relation between the diffusion coefficient and the real part of
the mobility:
$${D(\omega)\over\RE\mu(\omega)}=kT.\myeqno$$
Since $\mu(\omega)$ can be extended into an analytic function in the upper complex half plane ($\IM\omega>0$), Eq. (2.8) on
the one hand, and Eqs. (2.9) or (2.11) on the other hand, constitue two equivalent formulations of the first FDT.

Then, using the expression of $C_{vv}(\omega)$ as given by the first FDT (2.9), that is 
$$C_{vv}(\omega)={kT\over m}\,{2\RE\gamma(\omega)\over|\gamma(\omega)-i\omega|^2},\myeqno$$
one gets from Eq. (2.2) the noise spectral density,
$$C_{FF}(\omega)=mkT\,2\RE\gamma(\omega),\myeqno$$
a relation from which one deduces the expression of $\gamma(\omega)$:
$$\gamma(\omega)={1\over mkT}\int_0^\infty \bigl\langle F(t)F\bigr\rangle\,e^{i\omega t}\,dt.\myeqno$$
Eq. (2.14) is known as the second FDT [4],[5]. Eq. (2.13) on the one hand, and Eq. (2.14) on the other hand, are two
equivalent formulations of the second FDT. The second FDT can also be established directly by applying the linear response
theory to the force exerted by the bath on the particle, this force being considered as a dynamical variable of the isolated
particle-plus-bath system. We will come back to this point  in Section~3.

Thus, when the particle environment is a thermal bath, the two fluctuation-dissipation theorems are valid. In these two
theorems the bath temperature $T$ plays an essential role. Under its form (2.9) or (2.11), the first FDT involves
the spectral density of a dynamical variable linked to the particle (namely its velocity), while, under its form (2.13), the
second FDT involves the spectral density of the random force, which is a dynamical variable of the bath. 
\bigskip
{\bf 2.4. Regression theorem}

When the particle environment is a thermal bath, we may introduce in the  expression (2.6) of $\langle v(t)v(t')\rangle$ the
value of the ratio ${C_{FF}(\omega)/2\RE\gamma(\omega)}$ as given by the second FDT (2.13). One gets:
$$\bigl\langle v(t)v(t')\bigr\rangle={kT\over m}\,\int_{-\infty}^\infty{d\omega\over 2\pi}\,e^{-i\omega(t-t')}
\,\left\{{1\over\gamma(\omega)-i\omega}+{1\over\gamma^*(\omega)+i\omega}\right\}.\myeqno$$
In the above integral, the function ${1/[\gamma(\omega)-i\omega]}$ is analytic for $\IM\omega>0$, while the function
${1/[\gamma^*(\omega)+i\omega]}$ is analytic for $\IM\omega<0$ [5]. Computing
$\langle v(t)v(t')\rangle$ for
$t-t'>0$, one chooses as an integration contour the semi-circle of large radius in the lower complex half plane. The second
term does not contribute. One thus has:
$$\bigl\langle v(t)v(t')\bigr\rangle={kT\over m}\,\int_{-\infty}^\infty{d\omega\over 2\pi}\,e^{-i\omega(t-t')}
\,{1\over\gamma(\omega)-i\omega},\qquad t-t'>0.\myeqno$$
Formula (2.16) displays the fact that, when diffusion takes place in a thermal bath, the velocity correlation function is
characterized by the same law as the average velocity [8]. This result constitutes the regression theorem, valid at
equilibrium for any $\gamma(\omega)$. 

In particular, with the model for $\gamma(\omega)$ considered before ($\RE\gamma(\omega)
\propto|\omega|^{\delta-1}$ with $0<\delta<2$), one has:
$$\bigl\langle v(t)v(t')\bigr\rangle={kT\over m}\,E_{2-\delta}\bigl[-(\omega_\delta|t-t'|^{2-\delta}\bigr].\myeqno$$
In the particular case $t=t'$, one gets from Eq. (2.17) the equipartition result :
$$\bigl\langle v^2\bigr\rangle={kT\over m}.\myeqno$$

\mysection{Out of equilibrium environment}

The fully general situation of a particle diffusing in an out of equilibrium environment is much more difficult to describe.
Except for the particular case of a stationary environment, the motion of the diffusing particle cannot be
described by the generalized Langevin equation. A more general equation has to be used. The fluctuation-dissipation
theorems are {\sl a fortiori\/} not valid. However, one can try to extend these relations with the help of an
age and frequency dependent effective temperature, such as proposed and discussed for instance in [6],[7].

An equation of motion will be derived in the present Section for the case of a bilinear coupling between the particle and the
environment. As for the out of equilibrium extension of the linear response theory, it will be discussed in the next Section.
\bigskip
{\bf 3.1. Equation of motion of a particle linearly coupled to an out of equilibrium environment}

The validity of the generalized Langevin equation (2.1) is restricted to the case of a stationary medium. In other cases, for
instance when the diffusing particle evolves in an aging medium such as a glassy colloidal suspension of
Laponite [1]-[3], another equation of motion has to be used.  We shall derive such an equation by assuming a bilinear
coupling, of the form $-\Phi x$, between the particle and its environment. With such a coupling, the particle equation of
motion reads
$$m\,{dv\over dt}=\Phi(t),\myeqno$$
where $\Phi(t)$ denotes the global force exerted by the environment. In turn, the latter is perturbed by the coupling with the
particle. The linear response relation yielding the average force exerted by the environment is of the form
$$\bigl\langle\Phi(t)\bigr\rangle=\int_0^t\tilde\chi_{\Phi\Phi}(t,t')\,x(t')\,dt',\myeqno$$
where the causal function $\tilde\chi_{\phi\phi}(t,t')$ ($t>t'$) is a linear response function of the surrounding medium (in
Eq. (3.2), it is assumed that the particle and the environment have been put in contact at time $t=0$). The particle equation
of motion (3.1) takes the form
$$m\,{dv\over dt}=\int_0^t\tilde\chi_{\Phi\Phi}(t,t')\,x(t')\,dt'+\Phi_E(t),\myeqno$$
where $\Phi_E(t)$ is a fluctuating force of zero mean generated by the environment. When the latter is out of equilibrium,
there is no specific relation between the response function $\tilde\chi_{\Phi\Phi}(t,t')$ and the correlation function
$\langle\Phi_E(t)\Phi_E(t')\rangle$. In the most general situations, neither $\Phi_E(t)$ nor $v(t)$ can be viewed as stationary
random processes. Before considering these situations, let us come back briefly on the particular cases,
first, of a stationary medium, second, of a thermal bath, in order to see how the properties described in Section 2 can be
retrieved.
\bigskip
{\bf 3.2. Stationary medium case}

Let us consider again the particular case of a particle diffusing in a stationary medium, in order to see how the
generalized Langevin equation (2.1) can be deduced from the more general Eq. (3.3). When the medium is stationary,
$\tilde\chi_{\Phi\Phi}(t,t')$ reduces to a function of $t-t'$ ($\tilde\chi_{\Phi\Phi}(t,t')=\tilde\chi_{\Phi\Phi}(t-t')$).
Introducing then the causal function
$\tilde\gamma(t)$ as defined by
$$\tilde\chi_{\Phi\Phi}(t)=-m{d\tilde\gamma(t)\over dt},\myeqno$$
and integrating by parts, one gets from Eq. (3.3):
$$m\,{dv\over dt}=-m\int_0^t\tilde\gamma(t-t')\,v(t')\,dt'+\Phi_E(t)+m\tilde\gamma(0)x(t)-m\tilde\gamma(t)x(0).\myeqno$$
Let us now denote by $t_i$ (instead of $0$) the initial time at which the particle has been put in contact with its
environment. Eq. (3.5) then writes:
$$m\,{dv\over
dt}=-m\int_{t_i}^t\tilde\gamma(t-t')\,v(t')\,dt'+\Phi_E(t)+m\tilde\gamma(t_i)x(t-t_i)-m\tilde\gamma(t-t_i)x(t_i).\myeqno$$ Eq.
(3.6) can be identified with the generalized Langevin equation (2.1) by letting $t_i\to -\infty$ (one has then
$\tilde\gamma(t_i)=0$), and assuming that the term $-m\tilde\gamma(t-t_i)x(t_i)$, which depends on the initial position of the
particle, is negligible in this limit. The fluctuating force $\Phi_E(t)$ then identifies with the Langevin force $F(t)$.
The fluctuating force and the particle velocity can be viewed as stationary random processes. When equilibrium is not realized,
the fluctuation-dissipation relations (2.8) and (2.14) and the regression theorem (2.16) are not valid. 
\bigskip
{\bf 3.3. Thermal bath case}

When equilibrium is realized, the stationary medium constitutes a thermal bath. The linear response theory, applied to the 
bath dynamical variable $\Phi_E(t)$, yields
$$\tilde\chi_{\Phi\Phi}(t-t')=\beta\Theta(t-t'){\partial\over\partial
t'}\bigl\langle F(t)F(t')\bigr\rangle,\qquad\beta=(kT)^{-1},\myeqno$$
that is, using Eq. (3.4) and the causality property, and the fact that at equilibrium $\Phi_E(t)$ is identical with the
Langevin force
$F(t)$:
$$\tilde\gamma(t-t')={1\over mkT}\,\Theta(t-t')\,\bigl\langle F(t)F(t')\bigr\rangle.\myeqno$$
Eq. (3.8), through Fourier transformation, yields the second FDT (Eq. (2.14)).

\mysection{Out of equilibrium linear response theory}

When the environment is not stationary, response functions such as $\tilde\chi_{\Phi\Phi}(t,t')$ and 
$\tilde\chi_{vx}(t,t')$ (response function of the particle velocity) depend separately on the two times $t$ and $t'$ entering
into play, and not only on their difference
$\tau=t-t'$. However, the time difference
$\tau$, also called the measure time or the observation time, continues to play an essential role in the description. For this
reason, it has been proposed to define time and frequency dependent response functions as Fourier transforms with respect to
$\tau$ of the corresponding two-time quantities [6],[7],[11]. The time $t$, which represents the waiting time or the age of
the system, then plays the role of a parameter. 

Let us first briefly recall these definitions and examine in which conditions the age and frequency dependent response
functions share the analytic properties of the corresponding stationary quantities.
\bigskip
{\bf 4.1. Age and frequency dependent response functions}

Considering the response function $\tilde\chi(t,t')$ as a function of $t$ and of the measure time $\tau=t-t'$ ({\sl i.e.\/}
$\tilde\chi(t,t')=\tilde\chi_1(t,\tau)$), one introduces the Fourier transform of $\tilde\chi_1$ with respect to $\tau$ (for a
fixed $t$):
$$\chi_1(\omega,t)=\int\tilde\chi_1(t,\tau)\,e^{i\omega\tau}\,d\tau.\myeqno$$
Due to the causality of $\tilde\chi(t,t')$, the lower integration bound in formula (4.1) is $0$. As for the upper integration
bound, it is equal to $t$ minus the lower bound over $t'$. If one assumes that the perturbation is applied at $t'=0$, the
upper integration bound is equal to $t$. One then writes
$$\chi_1(\omega,t)=\int_0^t\tilde\chi_1(t,\tau)\,e^{i\omega\tau}\,d\tau,\myeqno$$
that is:
$$\chi_1(\omega,t)=\int_0^t\tilde\chi(t,t')\,e^{i\omega(t-t')}\,dt'.\myeqno$$
The Fourier relation (4.1) can be inverted, which yields:
$$\int_{-\infty}^\infty{d\omega\over
2\pi}\,e^{-i\omega(t-t')}\,\chi_1(\omega,t)=\Theta(t')\,\Theta(t-t')\,\tilde\chi(t,t').\myeqno$$ The time $t$ can be
interpreted as the waiting time or the age $t_w$ of the system under study. In other words, one has introduced [6],[7],[11]:
$$\chi_1(\omega,t_w)=\int_0^{t_w}\tilde\chi(t_w,t_w-\tau)\,e^{i\omega\tau}\,d\tau.\myeqno$$
Note that the stationary regime is obtained, not only by assuming that $\tilde\chi(t,t')$ is invariant by
time translation ($\tilde\chi(t,t')=\tilde\chi_1(\tau)$), but also, moreover, that the age of the system tends towards
infinity:
$$\chi(\omega)=\lim_{{t_w}\to\infty}\int_0^{t_w}\tilde\chi_1(\tau)\,e^{i\omega\tau}\,d\tau.\myeqno$$

However, from the  point of view of linear response theory, the definitions (4.1) or (4.5) suffer from several drawbacks.
Actually, the function $\chi_1(\omega,t_w)$ as defined by Eq. (4.5) is not the Fourier transform of the function
$\tilde\chi_1(t_w,\tau)$, but a partial Fourier transform computed in the restricted time interval $0<\tau<t_w$.\break As a
consequence, it does not possess the same analyticity properties as the generalized susceptibility $\chi(\omega)$ defined by
Eq.~(4.6). While the latter, extented to complex values of $\omega$, is analytic in the upper complex half plane
($\IM\omega>0$), the function $\chi_1(\omega,t_w)$ is analytic in the whole complex plane. As a very simple example, consider
the exponentially decreasing response function 
$$\tilde\chi(t,t')=\Theta(t-t')\,e^{-\gamma(t-t')}.\myeqno$$
One has\footnote{$^2$}{Despite of the time translational invariance of $\tilde\chi(t,t')$, the partial Fourier transform
$\chi_1(\omega,t_w)$ effectively depends on $t_w$, due to the fact that the perturbation is applied at time 0 and not
$-\infty$.}:
$$\chi_1(\omega,t_w)={1\over\gamma-i\omega}\,\bigl[1-e^{-(\gamma-i\omega)t_w}\bigr].\myeqno$$
The corresponding generalized susceptibility is obtained by taking the limit\break $t_w\to\infty$ in Eq. (4.8):
$$\chi(\omega)={1\over\gamma-i\omega}.\myeqno$$
In this example, $\chi(\omega)$ as given by Eq. (4.9) has a pole in the lower complex half plane ($\omega=-i\gamma$). This
pole can be traced back to an eliminable singularity in the r.h.s. of Eq. (4.8), so that $\chi_1(\omega,t_w)$ is analytic
everywhere. This is an important drawback, since the information about the nature of the modes of the unperturbed
system, which is contained in the poles of the generalized susceptibility, is not contained in the partial
Fourier transform $\chi_1(\omega,t_w)$ as defined by Eq. (4.5). 
\bigskip
{\bf 4.2. Quasi-stationary regime: introduction of an effective temperature}

However, if $\omega t_w\gg 1$, one can assume that $\chi_1(\omega,t_w)$ varies very slowly with $t_w$ in the range of values
of $\omega$ of interest [6],[7].  It exists then two well separated time scales, respectively
characterizing the times pertinent for the measuring process, and the waiting time or the age of the system. In this case, 
$\chi_1(\omega,t_w)$ describes a quasi-stationary regime taking place for a given waiting time. Considered as a function
of $\omega$, $\chi_1(\omega,t_w)$ is assumed to have the analytic properties required from a generalized susceptibility ({\sl
i.e.\/} analyticity in a complex half plane, namely the upper one with our definition of the Fourier transformation). From now
on, we will drop out the parameter $t_w$ for simplicity, being it understood that the analysis is valid for a given $t_w$.

Let us now come back to the specific problem of the diffusion of a particle in an out of equilibrium environment. In a
quasi-stationary regime, the particle velocity obeys the generalized Langevin equation (2.1) The generalized
susceptibilities of interest are the particle mobility $\mu(\omega)=\chi_{vx}(\omega)$ and the friction coefficient
$\gamma(\omega)=-({1/m\omega})\chi_{\Phi\Phi}(\omega)$ (this latter formula being a consequence of the relation (3.4) between
$\tilde\gamma(t)$ and $\tilde\chi_{\Phi\Phi}(t)$). The results of the linear response theory as applied to the particle
velocity, namely the first FDT (2.8) and the Einstein relation (2.11), are not valid out of equilibrium. One can then try to
extend the linear response theory with the help of an effective temperature [6],[7]. More precisely, we shall write the
relation between the velocity spectral density
$C_{vv}(\omega)=\int_{-\infty}^\infty\langle v(t)v\rangle e^{i\omega t}dt$ and the real part of the mobility as
$$C_{vv}(\omega)=kT_{\rm eff.}(\omega)\,2\RE\mu(\omega),\myeqno$$
which amounts to assume a modified Einstein relation:
$${D(\omega)\over\RE\mu(\omega)}=kT_{\rm eff.}(\omega).\myeqno$$
Eqs. (4.10) and (4.11) define a frequency dependent effective temperature $T_{\rm eff.}(\omega)$. Note,
however, that it is not possible to rewrite the first FDT under a form similar to Eq. (2.8) with $T_{\rm eff.}(\omega)$ as
defined by Eqs. (4.10) and (4.11) in place of $T$:
$$\mu(\omega)\neq{1\over kT_{\rm eff.}(\omega)}\int_0^\infty\bigl\langle v(t)v\bigr\rangle\,e^{i\omega t}\,dt.\myeqno$$

Nevertheless, $T_{\rm eff.}(\omega)$ can consistently be used in the relation between the noise spectral density
$C_{FF}(\omega)=\int_{-\infty}^\infty\langle F(t)F)\rangle e^{i\omega t}dt$ and the real part of the friction coefficient.
Indeed, from Eq. (2.2), valid in any stationary regime, and Eq.~(4.10), one deduces:
$$C_{FF}(\omega)=mkT_{\rm eff.}(\omega)\,2\RE\gamma(\omega).\myeqno$$
But, as for the first FDT, it is not possible to rewrite the second FDT under a form similar to Eq. (2.14) with $T_{\rm
eff.}(\omega)$ in place of $T$:
$$\gamma(\omega)\neq{1\over mkT_{\rm eff.}(\omega)}\int_0^\infty\bigl\langle F(t)F\bigr\rangle\,e^{i\omega t}\,dt.\myeqno$$

Given $\RE\gamma(\omega)$ and $T_{\rm eff.}(\omega)$, it is possible to compute by Fourier analysis the out of equilibrium
velocity correlation function $\langle v(t)v(t')\rangle$, and to deduce from it the associated mean square displacement $\Delta
x^2(t)$. The results can be compared with the ones obtained at equilibrium ($T_{\rm eff.}(\omega)=T$) for the same
$\RE\gamma(\omega)$ [8]. Such a program will be carried out in the following Section.

\mysection{Anomalous diffusion in an aging medium}

In terms of the effective temperature $T_{\rm eff.}(\omega)$ as defined by Eq. (4.10) and Eq.~(4.11), one has:
$$\bigl\langle v(t)v(t')\bigr\rangle=\int_{-\infty}^\infty{d\omega\over
2\pi}\,e^{-i\omega(t-t')}\,{kT_{\rm eff.}(\omega)\over
m}2\RE\mu(\omega).\myeqno$$
Formula (5.1) displays the fact that, when diffusion takes place in an aging medium as pictured by a frequency-dependent
effective temperature $T_{\rm eff.}(\omega)$, the velocity correlation function is not characterized by the same law as the
average velocity. In other words, the regression theorem (2.16) is not valid. Accordingly, the equipartition result (2.18) is
not valid either. 

To characterize at best the properties of $\langle v(t)v(t')\rangle$ in an aging medium, we shall take for $\RE\gamma(\omega)$
the same function than in our equilibrium study~[8], namely a function behaving like a power-law characterized by the exponent
$\delta-1$:
$$\RE\gamma(\omega)=\gamma_\delta\,\Bigl({|\omega|\over\tilde\omega}\Bigr)^{\delta-1},\qquad|\omega|\ll\omega_c.\myeqno$$
We limit ourselves to the case $0<\delta<2$, in which an initial fluctuation of the particle velocity relaxes towards zero at
large times, as displayed by Eq. (2.4).
In Eq. (5.2), $\omega_c$ denotes a cut-off frequency typical of the environment, and $\tilde\omega\ll\omega_c$ is a
reference frequency allowing for the coupling constant
$\eta_\delta=m\gamma_\delta$ to  have the dimension of a viscosity for any $\delta$ [12]. Since $\delta<2$, the real part of 
the mobility $\mu(\omega)={1/(m[\gamma(\omega)-i\omega])}$ varies like $\omega^{1-\delta}$ for
$\omega\ll\tilde\omega$:
$$\RE\mu(\omega)\sim{1\over
m\gamma_\delta}\,\Bigl({|\omega|\over\tilde\omega}\Bigr)^{1-\delta}\,\sin^2{\delta\pi\over
2},\qquad|\omega|\ll\tilde\omega.\myeqno$$

Provided that the behaviour of $T_{\rm eff.}(\omega)$ at small $\omega$ is known, the Fourier relation
(5.1) allows for a simple asymptotic analysis of the large time behaviour of the out of equilibrium velocity correlation
function.
\bigskip
{\bf 5.1. Modelization of the function $\bf T_{\bf eff.}(\omega)$}

Let us assume the following behaviour of $T_{\rm eff.}(\omega)$ at small $\omega$:
$$T_{\rm eff.}(\omega)\sim T\,\Bigl({|\omega|\over\omega_0}\Bigr)^\alpha,\qquad|\omega|\ll\omega_0,\qquad\alpha<0.\myeqno$$
In Eq. (5.4), both the exponent $\alpha$ and the characteristic frequency $\omega_0\ll\tilde\omega$ may depend on the waiting
time
$t_w$  ($\alpha=\alpha(t_w)$, $\omega_0=\omega_0(t_w)$). The frequency
$\omega_0$ is a decreasing function of $t_w$ [13],[14]. It separates low frequencies ({\sl i.e.\/} slow modes), for which
the modelization (5.4) applies, from high frequencies ({\sl i.e.\/} fast modes), for which one has:
$$T_{\rm eff.}(\omega)\sim T,\qquad|\omega|\gg\omega_0.\myeqno$$
The chosen modelization of the effective temperature in an aging medium insures that the density of slow
modes in the noise is proportionally larger than in a thermal bath at temperature $T$ [14]. Note that the equilibrium situation
would correspond to $\omega_0=0$, since then $T_{\rm eff.}(\omega)=T$ for any $\omega$. It can also formally be retrieved by
taking the limit $\alpha\to 0^-$ in the low frequency modelization~(5.4). 
\bigskip
{\bf 5.2. Asymptotic analysis}

Eq. (5.2), together with the low frequency expression (5.4) of $T_{\rm eff.}(\omega)$, shows that the velocity correlation
function is the Fourier transform of a function varying like $\omega^{\alpha+1-\delta}$ at small $\omega$ ({\sl i.e.\/}
$|\omega|\ll\omega_0\ll\tilde\omega$). The Fourier integral (5.1) converges provided that the condition
$$\alpha>\delta-2\myeqno$$
is fulfilled. According to the usual properties of the Fourier transformation, the correlation function $\langle
v(t)v(t')\rangle$ behaves like
$|t-t'|^{-\alpha+\delta-2}$ for large values of $|t-t'|$. More precisely, one has:
$$\bigl\langle v(t)v(t')\bigr\rangle\sim{kT\over
m}\,{\tilde\omega\over\gamma_\delta}\,\bigl(\omega_0|t-t'|\bigr)^{-\alpha}\,\bigl(\tilde\omega|t-t'|\bigr)^{\delta-2}\,{1\over
\Gamma(\delta-\alpha-1)}\,{\sin^2{\delta\pi\over 2}\over\sin{(\delta-\alpha)\pi\over 2}}.\myeqno$$
After two integrations over time (see formula (2.7)), one obtains the large time expression of the mean square displacement:
$$\Delta x^2(t)\sim 2\,{kT\over m\gamma_\delta}\,{1\over\tilde\omega}\,(\omega_0 t)^{-\alpha}\,(\tilde\omega
t)^\delta\,{1\over\Gamma(\delta-\alpha+1)}\,{\sin^2{\delta\pi\over 2}\over\sin{(\delta-\alpha)\pi\over 2}}.\myeqno$$

At this stage, it is convenient to introduce the $\alpha$- and $\delta$-dependent frequency $\omega_{\alpha,\delta}$ as defined
by
$$\omega_{\alpha,\delta}^{2-\delta+\alpha}=\gamma_\delta\,\omega_0^\alpha\,{1\over\tilde\omega^{\delta-1}}\,
{\sin{(\delta-\alpha)\pi\over 2}\over\sin^2{\delta\pi\over
2}}.\myeqno$$
In terms of $\omega_{\alpha,\delta}$, one has:
$$\Delta x^2(t)\sim 2\,{kT\over
m}\,{1\over\omega_{\alpha,\delta}^2}\,{(\omega_{\alpha,\delta}t)^{\delta-\alpha}\over\Gamma(\delta-\alpha+1)},\qquad\omega_{\alpha,\delta}t\gg
1.\myeqno$$ 
Out of equilibrium, the anomalous diffusion exponent, as defined by $\Delta x^2(t)\sim t^\nu$, is:
$$\nu=\delta-\alpha.\myeqno$$
Note that the condition (5.6) insures that $\nu<2$, as required since the motion must not be
kinematical. 

The equilibrium value $\nu_{\rm eq.}=\delta$ of the diffusion exponent is recovered, as it should, by making $\alpha=0$ in
Eq. (5.8). Moreover, the full equilibrium expression of the mean square displacement, that is [8] 
$$\Delta x^2(t)_{\rm eq.}\sim 2\,{kT\over m}\,{1\over\omega_\delta^2}\,{(\omega_\delta
t)^\delta\over\Gamma(\delta+1)},\qquad\omega_\delta t\gg 1,\myeqno$$
with the $\delta$-dependent frequency $\omega_\delta$ as defined by
$$\omega_\delta^{2-\delta}=\gamma_\delta\,{1\over\tilde\omega^{\delta-1}}\,{1\over\sin{\delta\pi\over 2}},\myeqno$$
is also retrieved from Eq. (5.7) in this limit, since 
$$\omega_\delta=\lim_{\alpha\to 0^-}\omega_{\alpha,\delta}.\myeqno$$.
\bigskip
{\bf 5.3. Discussion}

The value (5.11) of the anomalous diffusion exponent $\nu$ results from the frequency dependence of both the real part of the
friction coefficient and the effective temperature.

The out of equilibrium noise spectral density is given by Eq. (4.13), that is, with the real part of the friction
coefficient as given by Eq. (5.2) and the effective temperature as modelized by Eq. (5.4):
$$C_{FF}(\omega)=2mkT\gamma_\delta\,\Bigl({|\omega|\over\omega_0}\Bigr)^\alpha\,
\Bigl({|\omega|\over\tilde\omega}\Bigr)^{\delta-1},\qquad|\omega|\ll\tilde\omega_0\ll\tilde\omega.\myeqno$$
As indicated above, for a given $\RE\gamma(\omega)$, the negative value chosen for $\alpha$ signifies that, in the out of
equilibrium medium under study, the weight of slow modes in the noise is increased with respect to the value it would have in a
thermal bath at temperature $T$. Accordingly, the anomalous diffusion exponent is enhanced : $\nu>\nu_{\rm eq.}$. 

From the experimental side, the diffusion exponent $\nu$ can be deduced from mean square displacement measurements, while the
value of $\delta$ can be obtained independently from mobility measurements. From both types of results, the exponent $\alpha$
characterizing the low frequency dependence of $T_{\rm eff.}(\omega)$ and thus describing the density of
slow modes in the aging medium can be deduced.

\mysection{Conclusion}

We have studied the anomalous diffusion of a particle in an aging medium. In a quasi-stationary regime, the particle
velocity is governed by a generalized Langevin equation. The medium being out of equilibrium, the
fluctuation-dissipation theorems are not valid. Both the Einstein relation and the relation between the real part
$\RE\gamma(\omega)$ of the friction coefficient and the noise spectral density can be written in a modified way which involves
an age and frequency dependent effective temperature $T_{\rm eff.}(\omega)$. We have first carefully made precise the
conditions under which such an effective temperature can be used. Then, we have studied the velocity correlation
function and the mean square displacement by means of asymptotic Fourier analysis.

Two main parameters enter into play, $\delta$ and $\alpha$, which characterize the low frequency behaviours
of $\RE\gamma(\omega)$ and $T_{\rm eff.}(\omega)$, assumed to be respectively proportional to $|\omega|^{\delta-1}$
($0<\delta<2$) and to $|\omega|^\alpha$ ($\alpha<0$).

Our main new results are as follows. In an aging medium, the velocity correlation function is different from its
equilibrium form. The regression theorem is not valid. The anomalous diffusion exponent is given by $\nu=\delta-\alpha$. Since
$\alpha<0$, the diffusion exponent is enhanced as compared to the value it would have, with the same $\RE\gamma(\omega)$,  in a
thermal bath. This is due to the fact that the weight of slow modes in the noise is increased.

Several questions remain to be solved. In particular, it will be interesting to extend the present study so as to obtain a
complete analytic description of the velocity correlation function and of the mean square displacement at any
times, {\sl i.e.\/} not only in the asymptotic regime. We have previously solved this question at equilibrium, in which case
we have shown that Mittag-Leffler functions play a central role in this description [8]. The out of equilibrium extension is
left out for future work.

\vfill
\break
\noindent
{\smalltitle Acknowledgements}

We wish to thank B. Abou, J.-B. Fournier and F. Gallet for helpful discussions on several aspects of this problem.
\bigskip
\parindent=0pt
{\smalltitle References}
\bigskip
\baselineskip=12pt
\frenchspacing

1. A. Knaebel, M. Bellour, J.-P. Munch, V. Viasnoff, F. Lequeux and J.L. Harden, Europhys. Lett. {\bf 52}, 73 (2000).

2. B. Abou, D. Bonn and J. Meunier, Phys. Rev. E {\bf 64}, 021510 (2001).

3. L. Bellon and S. Ciliberto, Physica D {\bf 168-169}, 325 (2002).

4. R. Kubo, Rep. Prog. Phys. {\bf 29}, 255 (1966).

5. R. Kubo, M. Toda and N. Hashitsume, {\it Statistical physics
\uppercase\expandafter{\romannumeral 2} : nonequilibrium statistical mechanics\/}, Second edition, Springer-Verlag, Berlin,
1991.

6. L.F. Cugliandolo, J. Kurchan and L. Peliti, Phys. Rev. E {\bf 55}, 3898 (1997).

7. L.F. Cugliandolo and J. Kurchan, Physica A {\bf 263}, 242 (1999).

8. N. Pottier, Physica A {\bf 317}, 371 (2003).

9. A. Erd\'elyi, {\sl Higher Transcendental Functions\/}, Vol. 3, McGraw-Hill, New-York (1955).

10. R. Gorenflo and F. Mainardi, International Workshop on the Recent Advances in Applied Mathematics, State of
Kuwait, May 4 - 7, 1996. Proceedings, Kuwait University, Department of Mathematics and Computer Science, 193 (1996).

11. G.J.M. Koper and H.J. Hilhorst, Physica A {\bf 155}, 431 (1989).

12. U. Weiss, {\sl Quantum dissipative systems\/}, Second edition, World Scientific, Singapore, 1999.

13. L.F. Cugliandolo, Lecture Notes, Les Houches (July 2002), preprint cond-mat/0210312.

14. L. Buisson, S. Ciliberto and A. Garcimart\'\i n, preprint cond-mat/0306462.

\bye